\documentclass[aps,prb,twocolumn,groupedaddress]{revtex4}
\usepackage{graphicx}
\begin{document}
\newcommand{\GC}{{\it G-C}}
\newcommand{\AT}{{\it A-T}}
\newcommand{\HHH}{$\bf H_0$}
\newcommand{\SSS}{$\bf S_0$}
\newcommand{\HH}{$\bf H_1$}
\newcommand{\hh}{$\bf H_2$}
\newcommand{\UU}{$\bf U$}
\newcommand{\CC}{{\bf C}}
\newcommand{\CCC}{{\bf C}^{\dagger}}
\newcommand{\DD}{\mathbf{D}}
\newcommand{\GG}{\mathbf{G}}
\newcommand{\ggamma}{\mathbf{\Gamma}}
\newcommand{\sss}{\mathbf{\Sigma}}
\newcommand{\PRB}{{\em Phys. Rev. B }}
\newcommand{\PRL}{{\em Phys. Rev. Lett. }}
\newcommand{\APL}{{\em App. Phys. Lett. }}
\newcommand{\SCI}{{\em science }}
\newcommand{\NAT}{{\em Nature }(London) }
\newcommand{\JPCA}{{\em J. Phys. Chem. A}}
\newcommand{\JPCB}{{\em J. Phys. Chem. B}}
\newcommand{\NL}{{\em Nano Lett.}}
\newcommand{\CPL}{{\em Chem. Phys. Lett.}}
\newcommand{\JCP}{{\em J. Chem. Phys. }}
\newcommand{\ital}{{\it et. al.}}
% number in the upper right-hand corner of the title page in preprint mode.
% Use the 'preprint-numbers' class option to override journal defaults
%\preprint{}
%\topmargin 0.25in
\title{Inter-base Electronic Coupling for transport through DNA}
% address or URL should go in the {}'s for \email and \homepage.
\author{H. Mehrez}
\altaffiliation{also ELORET, Mail Stop 229-1}
\author{M. P. Anantram}
\altaffiliation{also UCSC, Mail Stop 229-1}
\affiliation{NASA Ames Research center, Moffett Field, CA 94035-1000, USA}

%Collaboration name if desired (requires use of superscript-address

\date{\today}
\begin{abstract}
We develop a new approach to derive single state tight binding
(SSTB) model for electron transport in the vicinity of
valence-conduction bands of poly(G)-poly(C) and poly(A)-poly(T) DNA. 
The SSTB parameters are derived from {\it first principles} and are used to
model charge transport through finite length DNA. 
We investigate the rigor of reducing the full DNA Hamiltonian to SSTB
model to represent charge transport in the vicinity of 
valence-conduction band. While the transmission coefficient spectrum is
preserved, its position shifts in energy. Thymine is poorly represented and
its peak height is substantially reduced. This is attributed to the
abstraction of the HOMO-LUMO coupling to other eigen-states in the nearest
neighbor DNA bases, and can be corrected within $2^{nd}$ order time
independent perturbation theory.
Inter-strand charge transport has also been analyzed and it is found
that hopping to the nearest neighbor in the
complementary strand is the most important process except in the
valence band of poly(G)-poly(C), where hopping to the second nearest
neighbor between $3'-$ends is the most dominant process. As a result,
transport between $3'-$ends and $5'-$ends in the vicinity of valence band of
poly{G}-poly{C} is asymmetric. 
%In the $3'G\rightarrow C3'$ configuration,
%current is carried through the {\it second nearest neighbor} DNA
%basis and it is $\sim100\%$ larger than current from $5'G\rightarrow C5'$.  
\end{abstract}

%73.63.-b Electronic transport in Nano scale materials and structures 
%73.63.Nm Quantum wires
%87.14.Gg DNA, RNA 
%87.15.Aa Theory and modeling; computer simulation  

%\keywords{}
\maketitle

\section{introduction}
The growing interest in DNA as a molecular
device\cite{diventra2003,endres2004} has led to a wide
range of experimental as well as theoretical work in the field. 
However, charge transport results through DNA are still controversial\cite{fink1999,porath2000,yoo2001,pablo2001,adessi2003,adessi2003a,mehrez2003,
felice2001,conwell2000, ye1999,bruinsma2000, berlin2000,voityuk2001,roche2003,barnett2001,gervasio2002,maragakis2002,artacho2003,zhu2004}. 
Experimentally it is found that DNA can either be a good conductor\cite{fink1999}, a
semi-conductor\cite{porath2000,yoo2001} or even an
insulator\cite{pablo2001}. Theoretically, while the exact charge transport
mechanism is not clear, band transport,\cite{porath2000,adessi2003,adessi2003a,mehrez2003} 
polaronic transport,\cite{yoo2001,conwell2000}
fluctuation facilitated charge migration,\cite{ye1999,bruinsma2000} and
multi-step hopping\cite{berlin2000,voityuk2001}
have been investigated.
%However, transport results through DNA are still controversial\cite{
%fink1999,
%felice2001,adessi2003,mehrez2003,
%conwell2000, ye1999,bruinsma2000, berlin2000,voityuk2001,roche2003,barnett2001,gervasio2002,maragakis2002,artacho2003}.
%Experimentally, it is found that
%DNA can either be a good conductor\cite{fink1999}, a
%semi-conductor\cite{porath2000,yoo2001} or even an
%insulator\cite{pablo2001}. Theoretically, the exact transport
%mechanism is not clear. Band-structure-like transport has been
%proposed\cite{porath2000,felice2001,adessi2003,adessi2003a,mehrez2003}. 
%Other scenarios such as polaronic\cite{yoo2001,conwell2000},
%fluctuation facilitated charge migration\cite{ye1999,bruinsma2000},
%multi-step charge transport via hopping\cite{berlin2000,voityuk2001}
%have also been suggested.  
%The scope for the wide range of interpretations arises because
The narrowing down of the precise charge transport mechanism has been 
difficult because the base pairs are weakly coupled due to the large 
inter-base separation of $\sim 3.4$ \AA. This translates into relatively
small hopping parameters for electrons between base pairs at the HOMO
and LUMO energy levels (less than 125 meV), which results in valence
and conduction bandwidths of less than 500 meV. As a result of the narrow
bandwidth, charge transport in DNA is easily modified by environment effects,
which include counter-ions,
impurities, defects as well as hydration. Recently, {\it ab
initio} 
%analysis have been performed to investigate the electronic
%properties of DNA at different ambient conditions\cite{endres2004,felice2001,adessi2003,adessi2003a,mehrez2003,voityuk2001,barnett2001,barnett2003,gervasio2002,maragakis2002,artacho2003}. 
calculations have focused on determining inter-base
coupling\cite{endres2004,voityuk2001,artacho2003} so that DNA can
be represented within a single-state tight binding (SSTB) model. However, these
derivations suffer from the
limitations that they either cannot be extended to describe interaction
between different bases\cite{artacho2003} or can only
find interactions between HOMO states (valence band) of DNA
bases\cite{voityuk2001}. Moreover, the rigor of reducing
the full Hamiltonian to the SSTB
model in describing electronic transport through finite
length DNA has not been addressed before.\\
The aim of this work is to develop a general
approach to obtain tight binding parameters to describe intra-strand
and inter-strand interactions from 
{\it first principles}. 
This formalism will be applied to poly(G)-poly(C) and poly(A)-poly(T) DNA. 

\section{methodology}
%{\bf Model:} 
We describe our model to determine the tight
binding parameters for poly(G)-poly(C) here, and follow an identical
approach  for poly(A)-poly(T).
Initially one GC base pair is constructed 
using the Nucleic Acid Builder (NAB) software package\cite{nab,arnott1976,amber94}. The backbone 
of the structure generated by NAB is replaced by 
a hydrogen atoms, and the position of these H atoms are relaxed at
the MP2 level. The optimized structure of the hydrogen terminated DNA
bases are shown in Fig.\ref{GC-transport}-a.
Using this configuration, double strand B-DNA
structures\cite{dna-structure} of  four, six and  eight
GC base pairs are constructed. The self consistent Hamiltonian \HHH~
of this structure is calculated using density functional theory, where the
 B3LYP density functional and $6-31G$ basis set are used.\cite{gaussian98}\\
To work in an orthogonal basis set, we initially transform \HHH~to
\HH~such that, 
\begin{equation}
{\mbox \HH}={\mbox \SSS}^{-1/2}{\mbox \HHH} {\mbox
\SSS}^{-1/2},
\end{equation}
where \SSS~ is the overlap matrix. 
%\begin{equation}
%{\mbox \HH}={\mbox \SSS}^{-1/2}{\mbox \HHH} {\mbox \SSS}^{-1/2}.
%\end{equation}
Every diagonal sub block of \HH, which corresponds to a DNA base, is
diagonalized and its eigenvectors are used to construct a
block diagonal matrix \UU. Following this, \HH~is transformed to,
%${\mbox \hh}={\mbox \UU}^{\dagger}{\mbox \HH}{\mbox \UU}$.
\begin{equation}
{\mbox \hh}={\mbox \UU}^{\dagger}{\mbox \HH}{\mbox \UU}.
\end{equation}
In this representation of \hh, diagonal elements correspond to the
localized energy levels of DNA bases and off diagonal blocks
correspond to inter-base interactions. Independent from the simulated
system size (four, six or eight base pairs), we find
the hopping parameters between HOMO and/or LUMO states beyond the {\it second} nearest
neighbor base to be insignificant, and so they are neglected. Hence the
Hamiltonian \hh~ is truncated and transformed to: 
\begin{eqnarray}
{\bf H}&=&\sum_{\stackrel{n_g}{i=1\rightarrow N_g}}
\epsilon_{n_g,i} \CCC_{n_g,i}\CC_{n_g,i}\nonumber
+\sum_{\stackrel{n_c}{i=1\rightarrow N_c}}
\epsilon_{n_c,i}\CCC_{n_c,i}\CC_{n_c,i}\nonumber\\
&+&\hspace*{-0.5cm}\sum_{\stackrel{<n_g,n_g^\prime>}{i,j=1\rightarrow N_g}}
t_{n_g,i;n_g^\prime,j}(\CCC_{n_g,i}\CC_{n_g^\prime,j}+c.c.) \nonumber\\ 
&+&\hspace*{-0.5cm}\sum_{\stackrel{<n_c,n_c^\prime>}{i,j=1\rightarrow N_c}}
t_{n_c,i;n_c^\prime,j}(\CCC_{n_c,i}\CC_{n_c^\prime,j}+c.c.) \nonumber\\    
&+&\hspace*{-0.5cm}\sum_{\stackrel{<n_g,n_c>}{i=1\rightarrow N_{g};\;\;j=1\rightarrow N_{c}}}
t_{n_g,i;n_c,j}(\CCC_{n_g,i}\CC_{n_c,j}+c.c.) \nonumber\\              
%&+&\sum_{\stackrel{<n_c,n_c'>}{i,j=1\rightarrow N_c}} 
%t_{n_c,i;,n_c',j}(\CCC_{n_c,i}\CC_{n_c',j}+c.c.)\nonumber\\
&+&\hspace*{-0.4cm}\sum_{\stackrel{\ll n_g,n_c\gg}{i=1\rightarrow N_{g};\;\;j=1\rightarrow N_{c}}}
t_{n_g,i;n_c,j}(\CCC_{n_g,i}\CC_{n_c,j}+c.c.) \mbox{ .}\label{eq-hamiltonian}
\end{eqnarray}
$\epsilon_{n_g,i}$ ($\epsilon_{n_c,i}$) is the $i^{th}$ on-site energy of base $n_g$ ($n_c$), where the subscripts $g$ and $c$ refer to guanine and cytosine respectively. 
$t_{n_g,i;n_g^\prime,j}$ ($t_{n_c,i;n_c^\prime,j}$) is the hopping parameter between energy levels $i$ and $j$ of base pairs $n_g$ and $n_g^\prime$ ($n_c$ and $n_c^\prime$) respectively.
$t_{n_g,i;n_c,j}$ is the inter-strand hopping parameter between energy levels $i$ and $j$ of base pairs $n_g$ and $n_c$.
$N_g$ and $N_c$ are the eigen-states in a single guanine and cytosine respectively.
$\CCC$ and $\CC$ are the creation and 
annihilation operators and $c.c.$ is the hermitian conjugate. 
$<...>$ and $\ll...\gg$ represent $1^{st}$
and $2^{nd}$ nearest neighbor interactions, respectively.\\
In finding the parameters of the Hamiltonian in 
Eq.~(\ref{eq-hamiltonian}), we have used the central two base pairs of the 
simulated system to minimize edge effects. The Hamiltonian in Eq.~(\ref{eq-hamiltonian})
is referred to as the {\it full DNA model} in the remainder of this paper.
\section{results}
\subsection{SSTB parameters}
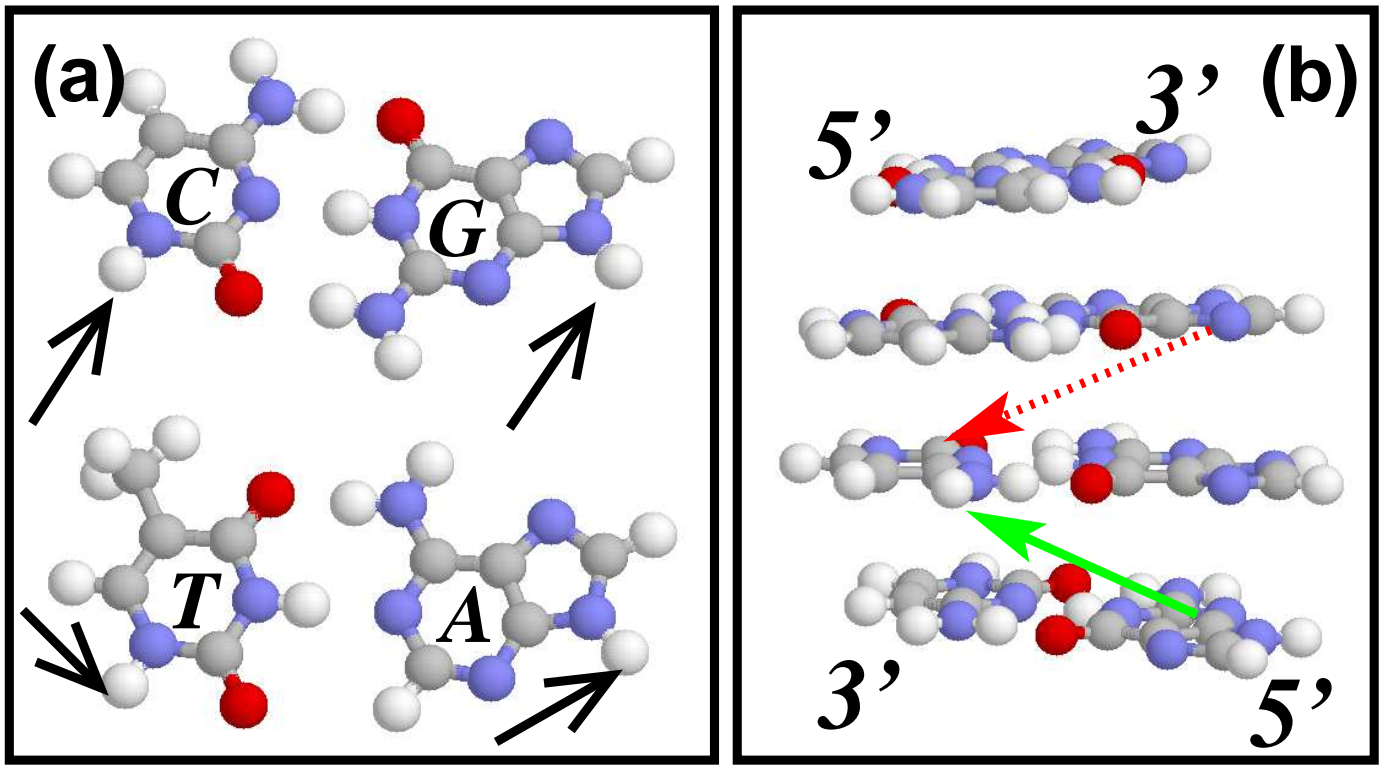
\begin{figure}
\caption{\label{GC-transport} 
(a) Atomic structure of hydrogen terminated DNA bases where arrows
indicate H atom which replaces the backbone. (b) 4 base pair
\GC~arranged in B-DNA 
configuration. $5',3'$-ends of DNA are shown and
the solid (dotted) arrow correspond to $<5'-G-C-5'>$ ($<3'-G-C-3'>$)
directional coupling.}
\end{figure}

We present our results for on-site energy and intra-strand hopping parameters of
the HOMO and LUMO states in table~\ref{intra-strand-coupling}. 
In poly(G)-poly(C), the HOMO (LUMO) state is localized on guanine (cytosine), 
and the hopping parameter between consecutive guanines (cytosines) is $115$ meV
($61$ meV). These values are much larger than the ones for poly(A)-poly(T), where the HOMO (LUMO) 
state is localized on adenine (thymine),
and the corresponding hopping parameter between consecutive adenines (thymines) is $21$ meV ($23$ meV). 
We have also determined the inter-strand hopping parameters which are 
shown in Table~\ref{inter-strand-coupling}.
Since the helical structure of the DNA breaks reflection symmetry in a plane perpendicular to the axis\cite{dna-structure}, 
we find that the inter-base hopping parameters depend on the directionality between $3'-$ and $5'-$ ends (Fig.\ref{GC-transport}-b). The most striking difference is  for the HOMO state of poly(G)-poly(C), where
$t_H \ll{\it 3'-G-C-3'}\gg=50$ meV (dotted line in
Fig.\ref{GC-transport}-b) and $t_H \ll{\it 5'-G-C-5'}\gg=7$ meV (continuous line in Fig\ref{GC-transport}-b).
We find that the directionality dependence of the inter-strand hopping parameters
cause significant asymmetry in inter-strand charge transport as discussed in
the following sub-section. Finally, we note that both the intra-strand and inter-strand hopping parameters do not depend on the
system size as indicated by the results for the four and eight
base pair systems shown in Tables~\ref{intra-strand-coupling} and ~\ref{inter-strand-coupling}.

\subsection{Transport Results}
%\subsection{Method}
%{\bf Transport Results:}
Charge transport experiments typically involve either measuring the
current-voltage characteristics of a DNA placed between metal
contacts\cite{fink1999} or measuring the charge transfer between donor
and acceptor intercalators placed along DNA strands.\cite{kelley1999}
While the parameters we derive here can be used to model both sets of
experiments, we will focus on the former experimental configuration in
the low bias limit. In presenting our results, 
we will compare the transmission probability in the vicinity of HOMO-LUMO states obtained using the 
full DNA  (Eq.~(\ref{eq-hamiltonian})) and SSTB
(Tables~\ref{intra-strand-coupling} and \ref{inter-strand-coupling})
models. However, to maintain the same injection rate from the leads to
the device in comparing the two models, we account for all eigen-states
of the edge base pairs that are coupled to the contacts. \\ 
The transport calculations are carried out within the
Landauer-B\"uttiker formalism\cite{landauer1957} and the transmission
probability is, 
\begin{equation}
T=tr[\Gamma^L {\bf G}^r \Gamma^R {\bf G}^a],
\end{equation}
where ${\bf G}^{r(a)}$ is the retarded (advanced) Green's function of
the isolated DNA attached to the contacts,
and $\Gamma^{L(R)}$ is the device coupling to the
left (right) contact. While the coupling, $\Gamma^{L(R)}$, is crucial
in determining the transport properties, it depends on the details of
DNA-contact coupling, which is difficult to control experimentally.    
%The retarded Green's function is obtained by solving,
%\begin{equation}
%[E - H - \Sigma^r_L(E) - \Sigma^r_R(E)] {\bf G}^r = I, \label{eq:gr}
%\end{equation}
%where E is energy, H is the Hamiltonian of the isolated DNA strand, and $\Sigma^r_L$ and $\Sigma^r_R$ are the retarded self energies representing
%coupling of the device to the left and right contacts
%respectively. In general,
%\begin{equation}
%\Gamma_{L(R)} (E) = - 2 Im[ \Sigma^r_{L(R)} (E) ],
%\end{equation}
%where Im stands for imaginary part.
%While the self energies that represent coupling of DNA to contacts is
%crucial in determining the transport properties, they depend on the details of DNA-contact
%coupling, which are difficult to control. 
We have considered two limits of the DNA-contact coupling:
the weak coupling limit where $\Gamma^L=\Gamma^R=10$~meV and the strong coupling limit where
$\Gamma^L=\Gamma^R=500$~meV \cite{model}. However, we present the results for
the latter configuration.\\ 
%The real part of the contact self energies are neglected,
%meaning that
%\begin{equation}
%\Sigma^r_{L(R)} = -i \Gamma_{L(R)} /2 \mbox{ ,} \label{eq:self}
%\end{equation}
%in Eq.~(\ref{eq:gr}).
Finally, we note that in modeling intra-strand transport both DNA strands
are coupled to the contacts at both ends. In the inter-strand
representation only one strand is coupled to the contact at each end.\\
%the same self energies given by Eq.~(\ref{eq:self}). 
%Inter-strand transport obviously involves coupling of only one strand to each contact.

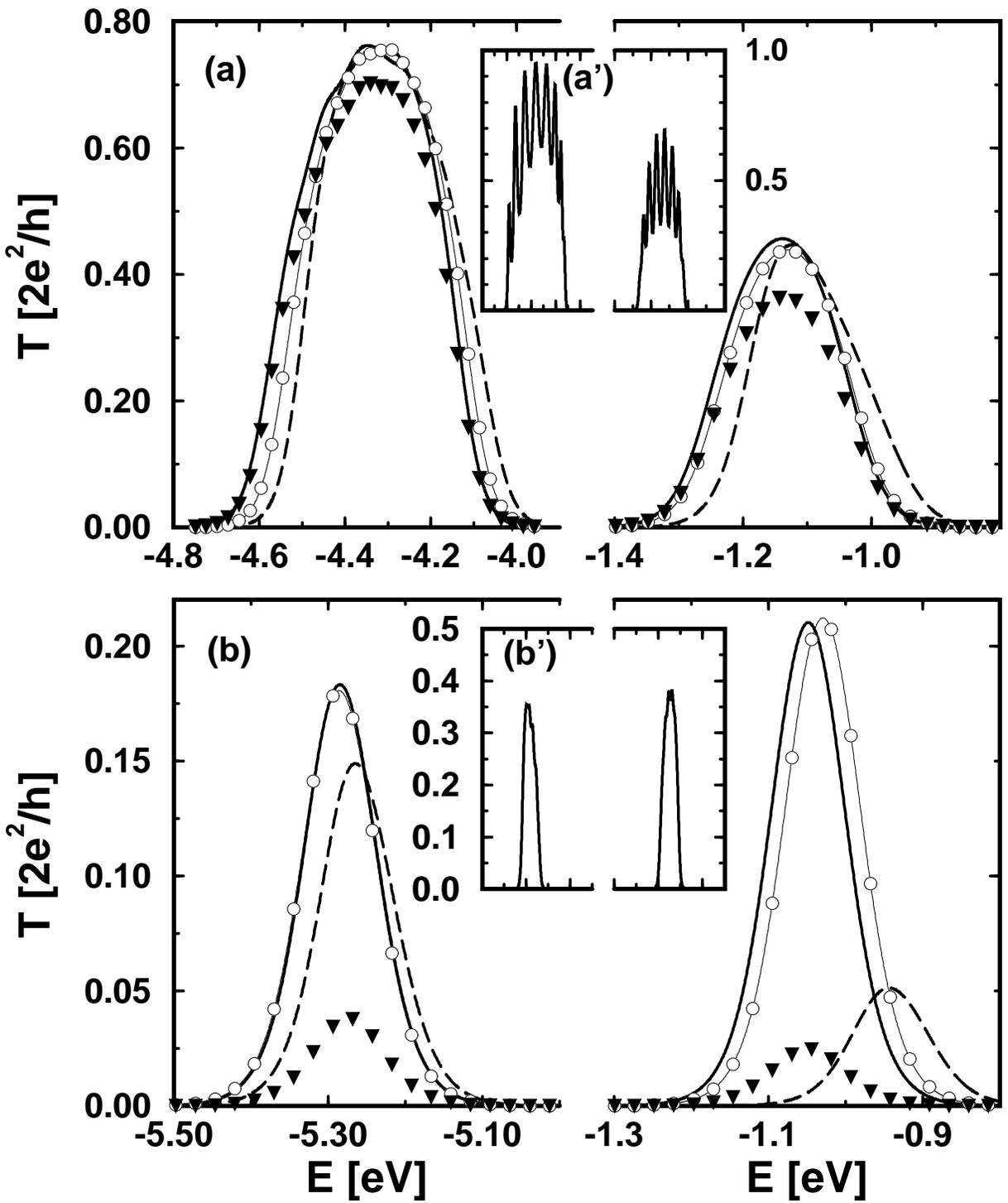
\begin{figure}
\caption{\label{GC-transport2} 
Intra-strand conductance of (a) poly(G)-poly(C) and (b) poly(A)-poly(T).
Left (Right) panels correspond to transport through HOMO (LUMO) states.
The linear response conductance is calculated at $300K$ using
$T(E)=\int dE'\,T(E')\,f(E-E')$ where $f$ is the Fermi function
evaluated at $300K$. 
Solid line - full dna model; Dashed lines - SSTB model with all eigen-states
at the two edges;  Open circles - Same as ``Dashed lines" but includes
the second order energy correction given in
Eq. (\ref{energy-correction}); Solid triangles 
- full DNA model minus the {\it tail effect}. Insets: Same as main
figures but conductance is calculated at $50 K$.}
\end{figure}

%\subsection{Results}
{\bf Intra-strand transport:}
The results for intra-strand transport from the full DNA and SSTB models are shown by the solid and dashed lines respectively, in 
Fig.~\ref{GC-transport2}. The SSTB model reproduces the peak and width
of the transmission windows quite well, with the main difference being
a shift in both the HOMO and LUMO transmission windows to higher
energies in the case of SSTB model. We find that this energy shift
occurs due to coupling of the HOMO and LUMO states of a base pair to
other eigen-states of neighboring base pairs. Further, because the
hopping parameters between the HOMO/LUMO states of a base pair and the other
eigen-states of neighboring base pairs is smaller than their energy
separation, we find that the energy shift seen in
Fig.~\ref{GC-transport2} can be quite accurately accounted for using
second order perturbation theory.   
%of eigen-states from the HOMO xxx
%, while calculations which
%incorporate a single state, either HOMO or LUMO, are presented as
%dashed lines. Fig.~\ref{GC-transport2}-a,b shows clearly that 
%transport characteristics through \GC~can be well described within 
%SSTB model.
%Better description of Guanine, compared
%to Cytosine, within single state model is attributed mainly to
%stronger inter-base coupling. Moreover, we note that 
%However, the spectrum of the transmission coefficient shifts to higher energy.
%This is due to the abstraction of the HOMO-LUMO
%coupling to other eigen-states in the nearest neighbor DNA bases. 
%Larger shift corresponds to stronger interaction and 
%suggests that Cytosine LUMO state is well coupled to other sates
%compared to Guanine HOMO state. Hence truncation of Cytosine basis is
%less accurate. We note that 
The expressions for the second order correction to the HOMO (H) or
LUMO (L) eigen-values are 
\begin{equation}
\Delta E^{(2)}_{n_g(n_c),H(L)}=-\sum_{\stackrel{\beta=n_g\pm
1,n_c\pm 1}{j\ne H(L)}} \frac{t^2_{n_g(n_c),H(L);\beta,j}}{\epsilon_{\beta,j}-
\epsilon_{n_g(n_c),H(L)} } \mbox{ .}\label{energy-correction}
\end{equation}
We find that the energy shift for G (C) bases at the center to be
$-49$ ($-69$) meV. The bases at the left and right 
edges have only one neighbor, and are shifted by 
$-20$ and $-29$ ($-29$ and $-40$) meV, respectively. When this shift
is included, the eigenvalues from the SSTB model match the full DNA
model. Calculations of the transmission probability which incorporate this second order 
correction in the SSTB model agree more closely with the full DNA model as shown by the open circles in Fig.~\ref{GC-transport2}-a.

The transmission probability for poly(A)-poly(T) are presented in Fig.~\ref{GC-transport2}-b. 
%Due to weak inter-base coupling,
%transmission width and peak height for transport along \AT is much
%smaller than the one corresponding to \GC. 
It is clear from the right panel of Fig.~\ref{GC-transport2}-b that the transmission
window of thymine  
undergoes both a large energy shift and substantial peak height
reduction compared 
to the case of poly(G)-poly(C). The reasons for this are: {\it (i)}
weak intra-strand 
hopping parameters for the LUMO level of thymine compared to cytosine (table \ref{intra-strand-coupling})
 and {\it (ii)} large second order energy correction to the LUMO state due to strong interaction
with other energy eigen-states. Following the perturbation theory analysis
of Eq.~(\ref{energy-correction}), we find the second order
energy correction to the LUMO eigen-state of thymine is $-100$
meV while that of the HOMO eigen-state of adenine is only $-41$ meV, for bases that are away from the edges.
The corresponding corrections for bases at the left and right edges are $-48$ and $-52$ ($-16$ and $-25$) meV respectively, for thymine (adenine). Again including these corrections in the Hamiltonian for the calculation of the transmission probability shows that the  SSTB 
model for poly(A)-poly(T) agrees closely to the full DNA model as shown by the open circles of
Fig.~\ref{GC-transport2}-b.\\ 

We now focus on the importance of including all eigen-states of the
edge base pairs connected to the contact, in modeling charge
transport. When the broadening due to the contacts $\Gamma^L$ and
$\Gamma^R$ is large, there is a non zero density of states in the
valence and conduction band energy windows at the edge base pairs due
to states other than $\epsilon_H$ and $\epsilon_L$. The hopping
parameter between these states at the edges and, $\epsilon_H$ and
$\epsilon_L$ of neighboring base pairs are non zero. As a result,
charge injected into the tails of energy eigen-states other than
$\epsilon_H$ and $\epsilon_L$, contributes to the transmission
probability at valence and conduction band energies. This contribution
is referred to as the {\it tail effect} and can 
%The contribution of the tail effect can 
be partitioned out of the
total transmission probability.
In Fig.~\ref{GC-transport2} we present transmission spectrum between only
HOMO (left panels of Fig.~\ref{GC-transport2}) and LUMO
(right panels of Fig.~\ref{GC-transport2}) states at the contact as triangle
symbols. Clearly, this corresponds to  $\sim 90\%$ and $ 25\%$ of
the total transmission for \GC~and \AT, respectively. It indicates
that contribution of other eigen-states to the total transmission
coefficient is important and they must be included by representing the
contact with all modes.\\
%and is shown as solid triangles in
%Fig.~\ref{GC-transport2}. Clearly, the tail effect has significant
%contribution to the total transmission probability in the valence
%(Fig.~\ref{GC-transport2}-a,c) and conduction
%(Fig.~\ref{GC-transport2}-b,d) bands. Their contributions are roughly
%$90\%$ and $25\%$ of the total transmission for poly(G)-poly(C) and
%poly(A)-poly(T) respectively.\\

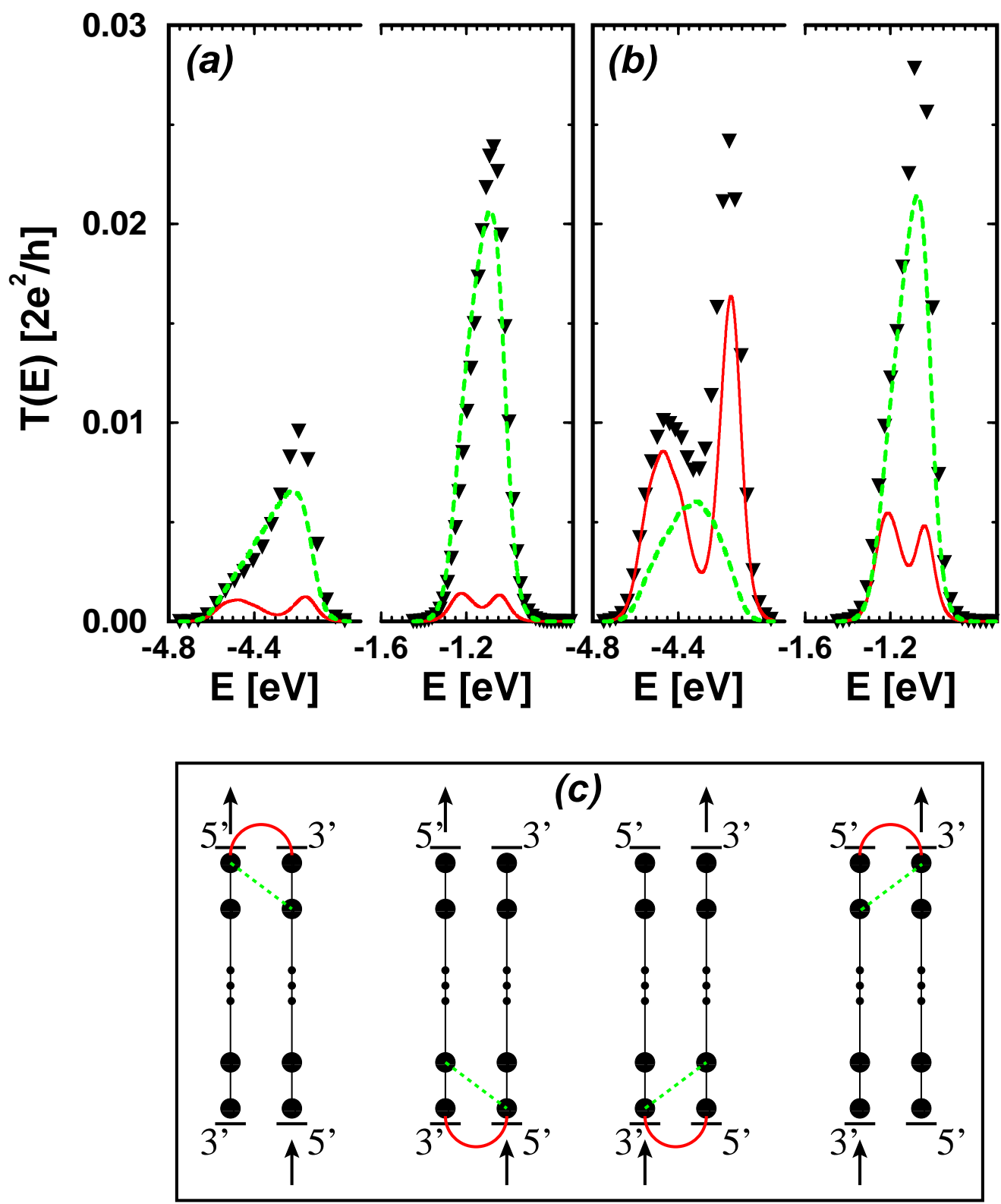
\begin{figure}
\caption{\label{inter-transport} Inter-strand conductance (calculated
at $300K$) for poly(G)-poly(C). Left (Right) panels
correspond to transport through HOMO (LUMO) states. In (a) only $5'-$ ends have an Ohmic contact
while in (b) only $3'-$ ends have an Ohmic contact.  Solid triangle - full model; Solid (Dashed) lines - Same as the solid triangle, except that the nearest (second nearest) neighbor shown by the solid (dashed) lines in (c) are set to zero. In (c), up-arrow ($\uparrow$) corresponds
to the Ohmic contacts where charge is  injected and collected. Each
diagram in (c) corresponds to the graph in (a) and (b) directly above it.
}
\end{figure}

{\bf Inter-strand transport:} Current-voltage measurements of {\it inter-strand} transport involve metal contacts connected to complementary strands.\cite{bhattacharya2003} That is, contacts are connected to either only $5'-$ends or
$3'-$ends of the DNA as shown in Fig.~\ref{GC-transport}-b. 
We first note that irrespective of which strand an electron is
injected into, transport occurs primarily along poly(G) (poly(C)) if
electrons are injected energetically into the valence (conduction)
band. Inter-strand hopping, which is the transmission limiting
step, occurs mainly near the contacts because of the tiny density of
states of the valence (conduction) band in poly(C) (poly(G)).  
We show the transmission probability for inter-strand transport in both the valence and conduction bands of poly(G)-poly(C) in Fig.~\ref{inter-transport}. 
The inter-strand transmission probability is more than an order of magnitude smaller than the intra-strand case because the density of states is peaked only along one strand in both the conduction and valence bands. Hopping into the strand with a smaller density of states limits the transmission / conductance in inter-strand transport.

We will now gain some insight into the roles of the nearest and second
nearest neighbor hopping parameters in table
\ref{inter-strand-coupling}, in determining inter-strand
transport. This is done by calculating the transmission probability by
setting specific inter-strand hopping parameters at the edges of the
DNA to zero in the Hamiltonian. The solid (dashed) lines in
Fig.~\ref{inter-transport}-a,b correspond to setting the nearest
(second nearest) inter-strand neighbor hopping parameter shown by the solid
(dashed) line respectively in Fig.~\ref{inter-transport}-c to
zero. The solid triangles in Fig.~\ref{inter-transport}-a,b is the
reference, which corresponds to the full model. 
Clearly, in the conduction band, the nearest neighbor inter-strand
hopping process shown by the solid line in
Fig.~\ref{inter-transport}-c is the most important in determining the
transmission probability between both the $3'-$ends and
$5'-$ends. Further, the conduction band transmission probabilities
across the  $3'-$ends and $5'-$ends are comparable. In contrast, in
the valence band, the transmission probability across the $3'-$ends is
twice as large as that across the $5'-$ends. The reason for this is
that the second nearest neighbor hopping parameter $t_H \ll{\it
  3'-G-C-3'}\gg$ determines the transmission probability across the
$3'-$ends, and is much larger (50 meV) than all other hopping
parameters in the valence band of poly(G)-poly(C) as seen in
Table~\ref{inter-strand-coupling}.\\
We have also investigated this asymmetric inter-strand transport for a
hundred base pair ($33.8$ nm long) poly(G)-poly(C) system. We have
found that it is is persistent and this represents a    
%We have also confirmed that this asymmetric inter-strand transport
%is persistent even for a hundred base pair long (~$33.8$ nm~)
%poly(G)-poly(C) system. This is a 
strong motivation to perform such inter-strand transport experiment. However,
we note that these results are sensitive to the DNA base
conformation. Hence thermal effects can substantially reduce the
asymmetric coupling along the DNA. Therefore, these experiments should
be performed at low temperature with ambient conditions that lead to
B-DNA conformation.\\ 
Inter-strand transport through poly(A)-poly(T) have also been carried
out and showed symmetric transmission between the $3'-$ends and
$5'-$ends. Inter-strand hopping occurs at nearest neighbor basis
closest to the contact region. 

\section{Conclusion}
%{\bf Conclusion:}

We have investigated the rigor of reducing the full Hamiltonian of a
DNA to a single-state tight binding (SSTB) model. Tight binding
parameters for both intra-strand and inter-strand transport have been
tabulated. We find that the SSTB model quite accurately reproduces the
transmission probability calculated from the full Hamiltonian, when
second order corrections to the on-site potential of base pairs is
included. As a result, the SSTB model is computationally efficient
when compared to the full Hamiltonian. One caveat while applying the
SSTB model is that injection of charge from the contacts to the edge
base pair should be carefully modeled. This is because charge can be
injected into the tails of all eigen-states of the edge base pairs,
thereby making conduction due to the {\it tail effect} important. 
We have also investigated inter-strand charge transport and found strong
asymmetric inter-strand current in the vicinity of the valence band
(HOMO state) of poly(G)-poly(C). Current flowing between Ohmic
contacts connecting the $3'$-ends is almost twice as large as the
current between Ohmic contacts connecting the $5'$-ends.\\
%
%{\bf{coefficient - probability,  interaction - hopping, base - basis - bases}}
%

%\section{Acknowledgments}
{\bf Acknowledgments}
HM was supported by NASA contract to UARC/ELORET and NSERC of Canada.
MPA was supported by NASA contract to UARC.

\newpage
\begin{table}[h]
\caption{\label{intra-strand-coupling} On-site energy and
intra-strand hopping parameters (meV) for HOMO (H) and LUMO (L) states. The parameters are
obtained from a four (eight) base pair system.} 

\begin{ruledtabular}
\begin{tabular}{|c c c c c|}
Base & $\epsilon_{H}$ &$\epsilon_{L}$& $t_L^{BB}$ & $t_H^{BB}$\\ \hline
G    &    -4278 (-4278)     &  1137 (1148)     &  19 (20)    & {\bf
  -115} (-114) \\ \hline
C    &    -6519 (-6533)    &  -1065 (-1072)     &  {\bf -61} (-60)   &  -24 (-21)\\ \hline
A    &    -5245 (-5245)    &  259  (258)     &  24 (25)   &    {\bf 21} (21)\\ \hline
T    &    -6298 (-6346)    &  -931 (-972)     &  {\bf -23} (-23)   &  -98 (-98)
\end{tabular}
\end{ruledtabular}
\end{table}

\begin{table}[h]
%\vspace*{0.2cm}\begin{table}[h]
\caption{\label{inter-strand-coupling} 
Inter-strand hopping parameters (meV) for HOMO (H) and LUMO (L) states. $<...>$ and
$\ll...\gg$ correspond to nearest neighbor and $2^{nd}$
nearest neighbor inter-strand interactions, respectively. The parameters are
obtained from a four (eight) base pair system.} 
\begin{ruledtabular}
\begin{tabular}{| lcr |}
System& $t_L$ & $t_H$\\ \hline
$<$G-C$>$         &63 (63)     &  2 (5)\\ \hline
$<$A-T$>$         &34 (34)     &  26 (26)\\ \hline
$\ll5'-G-C-5'\gg$ &-12 (-12)   & -7 (-8)\\ \hline
$\ll3'-G-C-3'\gg$ &-16 (-15) & 50 (48)   \\ \hline
$\ll5'-A-T-5'\gg$ &-10 (-10) & -11 (-11)\\ \hline
$\ll3'-A-T-3'\gg$ & -13 (-13) & 9 (9) 
\end{tabular}
\end{ruledtabular}
\end{table}

\newpage

%\begin{acknowledgments}
%\end{acknowledgments}


%\newpage
\bibliography{basename of .bib file}

\begin{thebibliography}{200}

\bibitem{diventra2003} 
M. Di Ventra and M. Zwolak, {\it DNA electronics}
in {\em Encyclopedia of Nanoscience and Nanotechnology}
edited by H. Singh-Nalwa, American Scientific Publishers (2004).

\bibitem{endres2004}
R. G. Endres, D. L. Cox, and R. R. P. Singh,
{\em Rev. Mod. Phys.}, {\bf 76}, 195 (2004).

\bibitem{fink1999}
H. W. Fink and C. Sch\"onenberger, 
\NAT {\bf 398}, 407 (1999);
%
\bibitem{yoo2001}
K.-H. Yoo, D. H. Ha, J.-O. Lee, J. W. Park, J. Kim, 
J. J. Kim, H.-Y. Lee, T. Kawai, and H. Y. Choi,
\PRL {\bf 87}, 198102 (2001);
%
\bibitem{porath2000}
D. Porath, A. Bezryadin, S. de Vries, and C. Dekker, 
\NAT {\bf 403}, 635 (2000);
%
\bibitem{pablo2001}
P. J. de Pablo, F. Moreno-Herrero, J. Colchero, J. G\'omez Herrero,
P. Herrero, A. M. Bar\'o, Pablo Ordej\'on, Jos\'e M. Soler, and Emilio Artacho
\PRL {\bf 85}, 4992 (2000).

\bibitem{felice2001}
R. Di Felice, A. Calzolari, E. Molinari, and A. Garbesi,
\PRB {\bf 65}, 045104 (2001).
 
\bibitem{adessi2003}
Ch. Adessi, S. Walch, and M. P. Anantram, 
\PRB {\bf 67}, R081405 (2003);
%Environment and structure influence on DNA conduction
\bibitem{adessi2003a}
Ch. Adessi and M. P. Anantram,
\APL {\bf 82}, 2353 (2003).
%Influence of counter-ion-induced disorder in DNA conduction

\bibitem{mehrez2003}
H. Mehrez, S. P. Walch and M. P. Anantram,
submitted to \PRB.
%Electronic properties of O$_2$ doped DNA.

\bibitem{conwell2000}
E. M. Conwell and S. V. Rakhmanova,
{\em Proc. Natl. Acad. Sci.} {\bf 97},  4556 (2000).
%Polarons in DNA

\bibitem{ye1999}
Y. J. Ye, R. S. Chen, A. Martinez, P. Otto, and
J. Ladik, 
{\em Solid St. Com.} {\bf 112}, 129 (1999). 

\bibitem{bruinsma2000}
R. Bruinsma, G. Gr\"uner, M. R. D'Orsogna, and J. Rudnick 
%Fluctuation-Facilitated Charge Migration along DNA
\PRL {\bf 85}, 4393 (2000).

\bibitem{berlin2000}
Y. A. Berlin, A. L. Burin, M. A. Ratner,
{\em J. Phys. Chem. A} {\bf 104}, 443 (2000).
%On the Long-Range Charge Transfer in DNA.

\bibitem{voityuk2001}
A. A. Voityuk, J. Jortner, M. Bixon, and N. R\"osch,
\JCP {\bf 114}, 5614 (2001);
%Electronic coupling between Watson-Crick pairs for hole transfer
%and transport in desoxyironucleic acid.
A. A. Voityuk, N. R\"osch, M. Bixon, and J. Jortner,
\JPCB {\bf 104}, 9740 (2000).

\bibitem{roche2003}
S. Roche, \PRL {\bf 91}, 108101 (2003); M. Unge and S. Stafstrvm, 
{\em Nano Lett.} {\bf 3}, 1417 (2003); 
O. R. Davis and J. E. Inglesfield, \PRB
{\bf 69}, 195110 (2004).

%D. M. Basko and E. M. Conwell,
%\PRL {\bf 88}, 098102 (2002).
%Effect of Solvation on Hole Motion in DNA

\bibitem{barnett2001}
R. N. Barnett,  C. L. Cleveland, A. Joy, U. Landman, G. B. Schuster,
{\em Science} {\bf 294}, 567 (2001).
%Charge Migration in DNA: Ion-Gated Transport
%Supplementary material is available at:
%www.sciencemag.org/cgi/content/full/294/5542/567/DC1

%\bibitem{barnett2003}
%R. N. Barnett, C. L. Cleveland, U. Landman, E. Boone, S. Kanvah, 
%and G. B. Schuster,
%\JPCA {\bf 107}, 3525 (2003).
%Effect of Base Sequence and Hydration on the Electronic and Hole 
%Transport Properties of Duplex DNA: Theory and Experiment

\bibitem{gervasio2002}
F. L. Gervasio, P. Carloni, and M. Parrinello,
\PRL {\bf 89}, 108102 (2002).
%Electronic Structure of Wet DNA

\bibitem{maragakis2002}
P. Maragakis, R. L. Barnett, E. Kaxiras,
M. Elstner, and T. Frauenheim,
\prb {\bf 66}, 241104(R) (2002).
%Electronic structure of overstretched DNA

\bibitem{artacho2003}
E. Artacho, M. Machado, D. S\'anchez-Portal,
P. Ordej\'on, and J. M. Soler,
{\em Molecular Physics} {\bf 101}, 1587 (2003).
%Electrons in dry DNA from density functional calculations

\bibitem{zhu2004}
Yu Zhu, Chao-Cheng Kaun, and Hong Guo 
%Contact, charging, and disorder effects on charge transport through a
%model DNA molecule
\PRB {\bf 69}, 245112 (2004)
\bibitem{nab}
NAB program (Version 4.5), T. Macke, W.A. Svrcek-Seiler, and D. A. 
Case.
In NAB program, the atomic position of B-DNA structure is 
%generated in
%the same way as NUCGEN which is part of AMBER (package for molecular
%simulaion). It 
based on X-ray diffraction data information to a resolution of $\sim
2$\AA, where heavy atoms are already tabulated in
Ref.[\onlinecite{arnott1976}]. Hydrogen atoms are added to the
system based on AMBER94 force field\cite{amber94}. 


\bibitem{arnott1976}
S. Arnott, P.J. Campbell-Smith, and R. Chandrasekaran.
In {\em Handbook of Biochemistry and Molecular Biology, 3rd
ed. Nucleic Acids} {\bf V. II}, 411, edited by G.P. Fasman, 
Ed. Cleveland: CRC Press, (1976).

\bibitem{amber94}
W.D. Cornell, {\it et. al.}, {\em J. Am. Chem. Soc.} {\bf 117}, 5179 (1995).
 
\bibitem{dna-structure}
In the B-DNA structure, base pairs repeat to form  a {\it one
dimensional} helix with an inter-base separation of $3.38$ \AA~and a
pitch angle of $36^o$. Such a pitch angle breaks symmetry along the
helix and defines a direction between the DNA ends. They are called
$5'-$ and $3'-$ ends of each strand. A {\it four} \GC~ base pair
structure with $5'-$ and $3'-$ ends description is shown in Fig.~\ref{GC-transport}-b.

%\bibitem{Lee2002}
%Hea-Yeon Lee, Hidekazu Tanaka, Yoichi Otsuka,
%Kyung-Hwa Yoo, Jeong-O Lee, and Tomoji Kawai,
%\APL {\bf 80}, 1670 (2002).
%Control of electrical conduction in DNA using oxygen hole doping



%\bibitem{becke1993}
%A. D. Becke,
%Density-functional thermochemistry. III. The role of exact exchange
%\JCP {\bf 98}, 5648 (1993);
%J. A. Pople, M. Head-Gordon, D. J. Fox, K. Raghavachari and L. A. Curtiss, 
%Gaussian-1 theory: A general procedure for
%prediction of molecular energies 
%\JCP {\bf 90}, 5622 (1989); 
%L. A. Curtiss, C. Jones, G. W. Trucks, K. Raghavachari and J. A. Pople, 
%Gaussian-1 theory of molecular energies for
%second-row compounds, 
%\JCP {\bf 93}, 2537 (1990).
 
\bibitem{gaussian98}
Gaussian 98 (Revision A.7), M. J. Frisch {\it et.al.}, Gaussian, Inc., 
Pittsburgh PA, 1998.
\bibitem{kelley1999}
S. O. Kelley and J. K. Barton, {\em Science} {\bf 283}, 375 (1999).

\bibitem{landauer1957}
R. Landauer,
{\em IBM J. Res. Dev.} {\bf 1}, 223 (1957);
%\PRB {\bf 16}, 4698 (1977);
%{\em IBM J. Res. Dev.} {\bf 32}, 306 (1988).
%\bibitem{buttiker1988}
M. B\"uttiker,
{\em IBM J. Res. Dev.} {\bf 32}, 317 (1988).
%\bibitem{datta_book}
S. Datta,
{\em Electronic Transport in Mesescopic Systems},
Cambridge University Press, New York, (1995).

%\bibitem{kelley1999}
%S. O. Kelley and J. K. Barton, {\em Science} {\bf 283}, 375 (1999).
\bibitem{model}
$\Gamma^{L,R}$ should be compared to $t^2_{DNA}/DoS_{DNA}(E_F)$ where
$t_{DNA}$ and $DoS_{DNA}(E_F)$ are the intra-strand hopping parameter and
DNA density of states at Fermi energy.
But for DNA
semi-infinite {\it one dimensional} system with coupling $t_{DNA}$,
$DoS_{DNA}(E_F)=1/\pi/t_{DNA}$. Hence, $\Gamma$ should be compared to
$t_{DNA}$.
\bibitem{bhattacharya2003}
S. Bhattacharya, J. Choi, S. Lodha, D. B. Janes, A. F. Bonilla,
K. J. Jeong, and G. U. Lee, 
IEEE NANO proceedings, 79 (2003).

\end{thebibliography}

\newpage

%\newpage
% Surround figure environment with turnpage environment for landscape
% figure

\end{document}